\documentclass{sigchi}
\toappear{\large{ACM ASSETS 2019 Workshop on \\AI Fairness for People with Disabilities}}
\usepackage{balance}
\usepackage{graphics}
\usepackage[T1]{fontenc}
\usepackage{txfonts}
\usepackage{mathptmx}
\usepackage[pdflang={en-US},pdftex]{hyperref}
\usepackage{color}
\usepackage{booktabs}
\usepackage{textcomp}
\usepackage{url}
\usepackage{microtype}
\usepackage{ccicons}

\def\doctitle{
Toward Fairness in AI for People with Disabilities: A Research Roadmap}
\def\plaintitle{
Toward Fairness in AI for People with Disabilities: \\A Research Roadmap
\vspace{-1pc}}
\def\plainauthor{Anhong Guo, Ece Kamar, Jennifer Wortman Vaughan, Hanna Wallach, Meredith Ringel Morris}
\def\plainkeywords{Artificial intelligence; machine learning; data; disability; accessibility; inclusion; AI fairness; AI bias; ethical AI.}

\makeatletter
\def\url@leostyle{
 \@ifundefined{selectfont}{
 \def\UrlFont{\sf}
 }{
 \def\UrlFont{\small\bf\ttfamily}
 }}
\makeatother
\def\pprw{8.5in}
\def\pprh{11in}

\definecolor{linkColor}{RGB}{6,125,233}
\hypersetup{
 pdftitle={\doctitle},
 pdfauthor={\plainauthor},
 pdfkeywords={\plainkeywords},
 pdfdisplaydoctitle=true,
 bookmarksnumbered,
 pdfstartview={FitH},
 colorlinks,
 citecolor=black,
 filecolor=black,
 linkcolor=black,
 urlcolor=linkColor,
 breaklinks=true,
 hypertexnames=false
}

\begin{document}

\title{\plaintitle}

\numberofauthors{1}
\author{
 \alignauthor{Anhong Guo$^{1,2}$, Ece Kamar$^1$, Jennifer Wortman Vaughan$^1$, \\ Hanna Wallach$^1$, Meredith Ringel Morris$^1$} \\ \smallskip
 \affaddr{$^1$ Microsoft Research, Redmond, WA \& New York, NY, USA}\\
 \affaddr{$^2$ Human-Computer Interaction Institute, Carnegie Mellon University, Pittsburgh, PA, USA}\\
 \email{ anhongg@cs.cmu.edu, \{eckamar, jenn, wallach, merrie\}@microsoft.com }
}

\maketitle

\begin{abstract}
AI technologies have the potential to dramatically impact the lives of people with disabilities (PWD). Indeed, improving the lives of PWD is a motivator for many state-of-the-art AI systems, such as automated speech recognition tools that can caption videos for people who are deaf and hard of hearing, or language prediction algorithms that can augment communication for people with speech or cognitive disabilities. However, widely deployed AI systems may not work properly for PWD, or worse, may actively discriminate against them. These considerations regarding fairness in AI for PWD have thus far received little attention. In this position paper, we identify potential areas of concern regarding how several AI technology categories may impact particular disability constituencies if care is not taken in their design, development, and testing. We intend for this risk assessment of how various classes of AI might interact with various classes of disability to provide a roadmap for future research that is needed to gather data, test these hypotheses, and build more inclusive algorithms. 

\end{abstract}



\section{Introduction}
As AI systems increasingly pervade modern life, ensuring that they work fairly for all is an important challenge. Researchers have identified unfair gender and racial bias in existing AI systems \cite{MachineBias, bolukbasi2016man, gendershades}. To understand how AI systems work across different groups of people, it is necessary to develop inclusive tools and practices for evaluation and to identify cases in which homogeneous, non-inclusive data \cite{gendershades} or data reflecting negative historical biases \cite{MachineBias, bolukbasi2016man} is used for system training.

Although improving the lives of people with disabilities (PWD)\footnote{Throughout this paper, we use people-first language as suggested by the ACM SIGACCESS guidelines \cite{a11ywriting}, but we recognize that some people may choose identity-first language or other terminology. Note that we use the term ``disability'' in accordance with the social model of disability \cite{oliver2013social}, which emphasizes that an impairment (i.e., due to a health condition or even a particular situational context) results in disability due to non-accommodating social or environmental conditions; under this model, AI systems could either mitigate or amplify disability depending on how they are designed.} is a motivator for many state-of-the-art AI systems, and although such systems have the potential to mitigate many disabling conditions \cite{earlyadopters}, considerations regarding fairness in AI for PWD have thus far received little attention \cite{trewin2018ai}. Fairness issues for PWD may be more difficult to remedy than fairness issues for other groups, particularly where people with particular classes of disability may represent a relatively small proportion of a population. Even if included in training and evaluation data, they may be overlooked as outliers by current AI techniques \cite{trewin2018ai}. Such issues threaten to lock PWD out of access to key technologies (e.g., if voice-activated smart speakers do not recognize input from people with speech disabilities), inadvertently amplify existing stereotypes against them (e.g., if a chatbot learns to mimic someone with a disability), or even actively endanger their safety (e.g., if self-driving cars are not trained to recognize pedestrians using wheelchairs).

We propose the following research agenda to identify and remedy shortcomings of AI systems for PWD: (1) Identify ways in which inclusion issues for PWD may impact AI systems; (2) Test inclusion hypotheses to understand failure scenarios and the extent to which existing bias mitigation techniques (e.g., \cite{feldman2015certifying, hardt2016equality, kamishima2012fairness}) work; (3) Create benchmark datasets to support replication and inclusion (and handle the complex ethical issues that creating such datasets for vulnerable groups might involve); and (4) Innovate new modeling, bias mitigation, and error measurement techniques in order to address any shortcomings of status quo methods with respect to PWD.

In this position paper, we take a step toward the first of these goals by reflecting on ways in which current key classes of AI systems may necessitate particular consideration with respect to different classes of disability. Systematically studying the extent to which these interactions exist in practice, or demonstrating that they definitely do not, is an important next step toward creating AI inclusive of PWD; however, articulating the extent of a problem is a necessary precursor to remediation. 

Furthermore, we note that the question of whether it is even ethical to build certain categories of AI is an important one (and may be dependent on use context). Our mention of various classes of AI is not an endorsement of whether we think such systems should be built, but is simply describing how they may interact with disability. Indeed, there is a larger ethical discussion to be had on how limiting some types of AI with negative associations (like synthetic voices that could be used for deepfakes \cite{deepfake}) might disenfranchise PWD who could benefit from such tech (i.e., by limiting the opportunity to realistically reproduce the voice of someone who can no longer speak). 

\section{Risk Assessment of Existing AI Systems for PWD}
Here, we group existing classes of AI systems by related functionalities, and identify disability constituencies for whom these systems may be problematic. This risk assessment is meant as a starting point to spark further research, and may not be exhaustive. For example, as new AI technologies are developed they would require consideration with respect to disability. Additionally, while we strove to anticipate ways in which classes of AI may fail for some disability groups, we may not have exhaustively identified all such groups; indeed, the ``long tail'' of disability and potential co-occurrence of multiple disabilities are two of many reasons that ensuring AI inclusion for PWD is particularly challenging \cite{trewin2018ai}. 

\subsection{Computer Vision}
Computer vision systems analyze still or video camera inputs to identify patterns, such as the presence and attributes of faces, bodies, or objects. Disabilities that may impact a person's physical appearance (facial features, facial expressions, body size or proportions, presence of assistive equipment, atypical motion properties) are important to consider when designing and testing the fairness of computer vision algorithms.

\subsubsection{Face Recognition}
Face recognition systems include capabilities for identifying the presence of a face and/or making inferences about its properties, including face \textit{detection}, \textit{identification} (i.e., to guess the identity of a specific person), \textit{verification} (i.e., to validate a claimed identity), and \textit{analysis} (e.g., gender classification, emotion analysis). Face recognition systems are already used in a wide variety of scenarios, including biometric authentication \cite{FaceID,WindowsHello}, security systems \cite{SmartGate}, criminal justice \cite{FBI-NGI}, interview support software \cite{HireVue}, and social/entertainment applications \cite{GooglePhotos}, many of which are controversial. 

We hypothesize that such techniques may not work well for people with differences in facial features and expressions if they were not considered when gathering training data and evaluating models. For instance, various aspects of facial analysis software may not work well for people with conditions such as Down syndrome, achondroplasia, cleft lip/palate, or other conditions that result in characteristic facial differences. Such systems may also fail for people who are blind, which may not only result in differences in eye anatomy, but may also result in a person wearing medical or cosmetic aids such as dark glasses, and may produce unanticipated behaviors, such as a person not holding their face toward a camera at the expected angle. Emotion processing algorithms may misinterpret the facial expressions of someone with autism or Williams syndrome, who may not emote in a conventional manner; expression interpretation may also be problematic for people who have experienced stroke, Parkinson's disease, Bell's Palsy, or other conditions that restrict facial movements.
 
\subsubsection{Body Recognition}
Body recognition systems include capabilities for identifying the presence of a body and/or making inferences about its properties, such as body detection, identification, verification, and analysis. Body recognition systems can power applications using gesture recognition (e.g., in VR and AR \cite{arkitbody, MRGestures} or gaming \cite{kinectbody}), or gait analysis (e.g., for biometric authentication \cite{wang2003silhouette}, sports biomechanics \cite{motionanalysis}, and path predictions used by self-driving vehicles \cite{umichpedestrian}).

Body recognition systems may not work well for PWD characterized by body shape, posture, or mobility differences. For example, gesture recognition systems\footnote{Many gesture systems use computer vision \cite{kim2012digits, MRGestures}, but some use other sensors, such as capacitive touchscreens \cite{elias2010multi}, accelerometers within devices \cite{guo2016tilt, laput2016viband}, etc.; body and mobility differences may create problems regardless of sensor type, though different sensor classes may have pros and cons for particular populations.} are unlikely to work well for people with differences in morphology (e.g., a person with an amputated arm may be unable to perform bimanual gestures, or may grip a device differently than expected; a person with polydactyly's style of touching a screen may register an unanticipated pattern). Failure of gesture recognition systems is also likely in cases where disability affects the nature of motion itself, such as for someone who experiences tremor or spastic motion \cite{mott2016smarttouch, mott2019clustertouch}. Fatigue may also impact gesture performance (and therefore recognition accuracy) over time, particularly for groups that may be more susceptible to fatigue such as due to disability or advanced age. The scheduling of medications whose main- or side-effects mitigate or amplify motor symptoms such as tremor may also result in differential gesture performance within or across days. 

People who are unable to move at all or who have severely restricted motion (e.g., people with ALS or quadriplegia), may be locked out of using certain technologies if body recognition is the only permitted interaction. Further, body recognition systems may not work well for people with mobility or morphology differences; for example, if a self-driving car's pedestrian-detection algorithm does not include examples of people with posture differences such as due to cerebral palsy, Parkinson's disease, advanced age, or who use wheelchairs during its training and evaluation, it may not correctly identify such people as objects to avoid, or may incorrectly estimate the speed and trajectory of those who move differently than expected, similar to Uber's recent self-driving car accident that killed a pedestrian walking a bicycle \cite{uberaccident}.

\subsubsection{Object, Scene, and Text Recognition}
Object, scene, and optical character recognition (OCR) systems recognize common objects, logos, text, handwriting, etc., and output labels, captions, and/or properties (i.e., location, activity, relationship). Systems taking advantage of these capabilities have been widely adopted by PWD, particularly people with visual impairments, such as Microsoft SeeingAI \cite{seeingai}, Google Lookout \cite{lookout}, LookTel Money Reader \cite{moneyreader}, KNFB Reader \cite{knfbreader}, OrCam MyEye \cite{orcam}, etc. 

Most systems for recognizing objects from photos are trained using datasets of images taken by people who are sighted, and the images are often of high quality since many are taken from social media sites such as Flickr \cite{deng2009imagenet}. These image data are known to be biased with regard to geographic areas and household income \cite{objectdetectionbias}. When applying the models to process images a blind user captures, the error rates often increase because images taken by people who are blind differ substantially in quality from those taken by people who are sighted due to poor framing, blur, unusual angles, poor lighting, etc. \cite{vizwiz-dataset}. Similar problems may be observed from pictures taken by people with tremor or other motor disabilities \cite{mott2018photography}. Further, OCR models for handwriting recognition may not work well for people with tremor or other motor disabilities that impact writing neatness. Additionally, error metrics used to evaluate many vision systems may not be adequate to capture the end-user experience of such tools, particularly for end users with disabilities that may prevent them from verifying the system's output (i.e., someone who is blind must rely on the output of an object detection system) \cite{macleod2017caption}. 

\subsection{Speech Systems}
We use the term ``speech systems'' to refer to AI systems that recognize the content (i.e., words) and/or properties (i.e., prosody, speaker demographics) of speech, or that generate speech from symbolic inputs such as text, Speech Synthesis Markup Language (SSML), or other encodings. Disabilities that may impact the content or clarity of a user's speech, as well as those impacting the ability to perceive sound, may reduce the accuracy and usability of speech systems. 

\subsubsection{Speech Recognition}
Automatic Speech Recognition (ASR) systems take in speech and output text. ASR systems have the potential to be important accessibility tools for people who are deaf or hard of hearing (DHH), such as by producing captions that can be overlaid as subtitles on videos \cite{GoogleAutomaticCaptioning,LiveCaption}, or possibly even using augmented reality to live-caption face-to-face speech \cite{ARCaptioning}. Speech input is also also useful for people who have difficulty using their hands to control traditional input devices \cite{AppleVoiceControl}.

ASR may not work correctly for people with atypical speech. ASR systems are known to have bias; for instance, many systems perform better for men than women \cite{nicol2002children, rodger2004field, tatman2017gender}. Today, many ASR systems do not work well for some older adults, due to differences in pitch, pacing, and clarity of speech by people of very advanced ages, since they are not commonly represented in the training and evaluation of the systems \cite{schlogl2013seniors}. People with accents, including accents due to disability (e.g., ``deaf accent''), also face challenges using current ASR tools \cite{fok2018towards, gottermeier2016user, tatman2017gender}, though it is possible to train personalized models for such groups \cite{GoogleSpeechImpairment, PowerPointCaption}. Speech disabilities such as disarthrya, as well as the use of speech-generating augmentative and alternative communication (AAC) devices, can also negatively impact ASR functionality \cite{kane2017aac}. Further, people who are unable to speak at all (i.e., some people who are deaf, people with some forms of aphasia), may be locked out of using ASR technologies. Additionally, error metrics used to evaluate many ASR systems, such as Word Error Rate, may not be adequate to capture the end-user experience of such tools, particularly for users with disabilities that may prevent them from verifying the system's output (i.e., someone who is profoundly deaf must trust the output of ASR captioning).

\subsubsection{Speech Generation}
Speech generation technologies include technologies such as text to speech (TTS) systems that aim to generate realistic audio from symbolic inputs such as text, SSML, or other markup, as well as emerging AI tools such as voice fonts \cite{chu2010providing, voicefont}, which aim to realistically mimic the sound of a particular speaker. TTS systems have been widely deployed in voice assistants such as Cortana, Alexa, Siri, and the Google Assistant; TTS is also key to many assistive technologies, including screen readers used by people who are blind and AAC devices used by people with speech and motor disabilities. Voice banking to create personalized voice fonts may be particularly valued by people with degenerative conditions that result in progressive loss of speaking abilities (e.g., ALS) \cite{fiannaca2018voicesetting,kane2017aac}. 

System defaults for what constitutes comprehensible speaking rates may need adjustments for particular disability segments; development of error metrics related to comprehension may need inclusion of such populations in order to account for diverse user needs -- for instance, people with cognitive or intellectual disabilities may require slower speech rates, whereas people with visual impairments may find rates too slow \cite{vtyurina2019verse}. Text-based prediction techniques are often deeply intertwined with speech generation in the case of AAC technologies; the choice of training and evaluation corpora for prediction may need to be adapted to be relevant to the topical needs and desired speech attributes of AAC users, supporting expressivity and authentic self-representation \cite{kane2017aac}.

\subsubsection{Speaker Analysis}
Speaker analysis systems include capabilities for speaker identification, speaker verification, and making inferences about the speaker's attributes such as age, gender, and emotion. Speaker analysis systems have a wide range of applications including biometric authentication \cite{nuance-voice-biometric}, enhancing speech transcription \cite{tranter2006overview}, and personalization \cite{google-assistant-multiuser}. Speaker analysis systems also have the potential to be important accessibility tools for people who are DHH, such as by supporting sound awareness through visualizations \cite{jain2015dhh}.

Speaker recognition and speech analysis tools that make inferences about a user's personal characteristics (i.e., gender, age) may not work well for PWD that significantly impact the sound of speech (e.g., dysarthria). Analysis tools that attempt to infer emotional state from prosodic features are likely to fail for speakers with atypical prosody, such as people with autism or some types of dementia.

\subsection{Text Processing}
Text processing systems perform functions related to understanding the content of text data, including tasks such as text analysis and translation. Text processing systems are likely to have accuracy and fairness challenges for people with cognitive and/or intellectual disabilities; systems for minority languages used by disability subcommunities, such as American Sign Language, are also a concern \cite{bragg2019sign}.

\subsubsection{Text Analysis}
Text analysis systems take text as input, and may attempt to detect content properties (e.g., key phrases, named entities, language) and/or author properties (e.g., sentiment, personality, demographics). Text analysis is broadly applied in record management, information retrieval, and pattern mining. Text analysis systems have the potential to be helpful for PWD that impact reading and writing, such as dyslexia, dysgraphia, or other cognitive differences, such as through visual illustration and focused highlighting \cite{immersivereader} or through intelligent spelling, grammar correction, and word or phrase suggestions \cite{grammarly}.

Cognitive and intellectual disabilities are likely to impact the efficacy and utility of many aspects of text analysis systems. For example, there is some evidence that spelling correction and query rewriting tools may not accurately handle dyslexic spelling \cite{morris2018dyslexia, rello2015spellchecker}. Further, people with autism may express emotion differently in writing than people who are neurotypical, resulting in incorrect classifications about their emotional state or personality. If these metrics are used as input to an automatic hiring system \cite{personalitytests} or automatic essay grading systems used with many standardized aptitude tests, text analysis systems can have accuracy and fairness challenges for people with cognitive and/or intellectual disabilities.

\subsection{Integrative AI}
In addition to the aforementioned classes of systems for vision, speech, and text processing, which were focused on single models, many complex AI systems are architectures integrating several models together to achieve more complex behavior. Here, we discuss two common examples of integrative AI: Information Retrieval and Conversational Agents.

\subsubsection{Information Retrieval}
Information retrieval (IR) tools, such as those that power web search engines, rely on AI for a variety of purposes, including query rewriting, autocompletion suggestions, spelling corrections, search result ranking, content summarization, and question answering. The input and output of IR systems can have many formats, e.g., image, video, sound, or text.

It is likely that many IR systems may inadvertently amplify existing biases against PWD, such as through returning stereotyped and/or over- and under-represented content in search results (a problem that has been documented with respect to gender in image search results \cite{kay2015unequal} and word embeddings \cite{bolukbasi2016man}). AI systems for advertising, both \textit{content-based} (i.e., related to the current search query) and \textit{behavior-based} (i.e., related to a user's personal characteristics), are also a key component of many commercial IR systems, as well as other online ecosystems (e.g., social media). Advertising algorithms and other types of recommender systems may hold particular risk for PWD by actively propagating discriminatory behavior such as through differential pricing for products and services and/or differential exposure to employment or other opportunities (an issue for which Facebook recently encountered legal trouble, by allowing housing ads that may have differentiated among protected demographics, including PWD \cite{FacebookHousingDiscrimination}). IR systems may pose particular challenges for people with cognitive or intellectual disabilities if not trained and tested with these groups; for example, people with dyslexia have reported that status quo query completion and result ranking techniques may not match their abilities \cite{morris2018dyslexia}.

\subsubsection{Conversational Agents}
Conversational agents provide conversational experiences to end users for various practical applications, including customer service \cite{facebookchatbots}, education \cite{duolingochatbot}, and health support \cite{fulmer2018using}. They are also powered by a variety of models, e.g., ASR, text analysis, TTS, and/or speaker analysis. Conversational agents have the potential to reduce users' workload when completing unfamiliar tasks \cite{guo2019statelens}, and could potentially provide cognitive assistance to people with dementia or intellectual disabilities that impact memory or executive functioning \cite{lewis2005hci}. 

If not carefully built, conversational agents could amplify existing biases against PWD, such as through returning stereotyped content in conversations (e.g., Microsoft shut down the chatbot Tay because it started generating hate speech learned from coordinated malicious users \cite{taychatbot}). Further, conversational agents may not work well for people with cognitive and/or intellectual disabilities, resulting in poor user experience. Training conversational agents on corpora that include data from people with a variety of cognitive and intellectual capabilities, as well as testing with similarly diverse audiences, is particularly important. For example, conversational agents may need to correctly interpret atypical spelling or phrasing from users with dyslexia, or may need to adjust their vocabulary level to be understood by someone with dementia. Further, conversational agents may need to support conversation in a user's preferred expressive medium, which may not be written language for some disability segments -- i.e., it may be important to support communication via sign languages (for people who are deaf) or via pictures and/or icons (for people with aphasia or autism).

\subsection{Other AI Techniques}
In addition to assessing risk factors for particular classes of AI applications, it is also worth considering that many AI techniques and practices that comprise the building blocks of such systems may lead to biases against PWD, such as techniques for outlier detection, practices of evaluating systems through aggregate metrics, definition of objective functions, and using training data that do not capture the true use cases or the true complexity of the real world.

Outlier detection algorithms flag outlier input, typically for punitive action, such as fraud detection. Lack of or low representation in training and evaluation data may erroneously result in people with a variety of disabilities being inadvertently flagged by anomaly detection tools, even when their actions should constitute legitimate system inputs. For example, many systems use task completion time as a signal for automatically determining input legitimacy, ranging from CAPTCHAs that aim to distinguish humans from bots to online crowd labor markets that aim to distinguish legitimate workers from spammers \cite{zyskowski2015accessiblecrowdwork}. However, many types of disability might manifest in atypical task performance timing, including the use of screen reader or magnifier tools by people with vision impairments, difficulty performing quick and accurate motions by people with a variety of motor-limiting conditions, people accessing devices through switch inputs due to motor limitations, slow reading times due to cognitive disabilities such as dyslexia, etc.

A common approach in evaluating AI systems is measuring performance with aggregate metrics such as accuracy, area under the curve (AUC), or mean square error (MSE). Aggregate metrics hide how performance varies across groups, in particular performance drops for small classes such as PWD \cite{nushi2018towards}. Objective functions that aim to maximize aggregate metrics will likely fail to prioritize performance for PWD. Recent work has introduced techniques that expand the objective functions for model training with terms that penalize performance discrepancies between subgroups \cite{agarwal2018reductions}.

Most AI systems are trained with existing datasets (i.e., data scraped from public corpora such as Flickr images \cite{deng2009imagenet}). In some cases, existing data sets may fail to capture the complexity of the real world and may lack representation of diverse groups, such as PWD. This may lead to blind spots in AI models \cite{lakkaraju2017identifying}. Actively curating inclusive datasets may be particularly important not only for training, but also for testing AI systems against known benchmarks.
 
\section{Discussion}
Our research roadmap for increasing fairness in AI for PWD included four proposed steps; this position paper mostly focused on the first: identifying ways in which (lack of) inclusion in training and evaluation of AI systems may negatively impact such systems' fairness for PWD. To address this, we discussed ways in which common categories of AI may need to account for various types of disabilities.

Regarding the types of potential harm caused by unfair AI, most of our examples are related to \textit{quality of service} \cite{gendershades}, like voice-activated smart speakers that may not recognize input from people with speech disabilities. Others are related to \textit{harms of allocation} \cite{MachineBias}, like using an incorrect prediction of the emotional state or personality of someone with autism as input into an automatic hiring system, or \textit{denigration} \cite{taychatbot}, like erroneously flagging inputs from PWD as invalid outliers. Additional potential harms include \textit{stereotyping} \cite{bolukbasi2016man} and \textit{over- or under-representation} \cite{kay2015unequal}; IR systems may inadvertently amplify existing biases against PWD by returning stereotyped and/or poorly represented content in search results. For issues related to allocation, quality of service, and representation, measuring objective fairness metrics through benchmarking could be sufficient to reveal bias, while issues related to stereotyping and denigration might require additional qualitative investigations. More thorough considerations of all types of harms with regard to PWD is important for future work.

In some cases, as indicated by the referenced citations, evidence already exists of problems for certain classes of AI for certain disability groups. For others, we have proposed hypotheses based on our knowledge of the domain space and analogous error cases for other minority user groups; our use of cautionary language such as ``may cause'' or ``is likely'' reflects this uncertainty. Per point \#2 of our research roadmap, systematic testing of the hypotheses we have presented here is a necessary step for future research. 

Item \#3 on our roadmap is the creation of public datasets for testing and benchmarking (and handling the complex ethical issues that creating such datasets for vulnerable user groups might involve); this is another key area for future work. Questions that must be addressed include: Is it acceptable to create datasets representing disability by scraping existing online data sources? How could this be done in a way that preserves users' privacy, and that ensures accurate ground-truth labeling of disability status? Are there potential harms that aggregating data about disability might expose people to? Could such data adequately cover rare conditions and/or intersectionality (either of co-occuring disabilities or of the intersection of disability with other demographics such as gender, race, geolocation, or socioeconomic status)? If curating data from scratch, how can we encourage contributions from target constituencies, and how can we ensure data collection mechanisms are sufficiently accessible? Is informed consent in data collection possible for people with some types of intellectual disabilities, and, if not, what methods can be used to promote fairness in AI for this class of end-user? Will it ever be possible to know one has complete coverage of all relevant disability communities, or will this always be an ``unknown unknown'' \cite{lakkaraju2017identifying}?

If, as we suspect may be the case, status quo modeling, bias mitigation, and/or error measurement techniques are inadequate for many scenarios affecting PWD, further research into new techniques will be warranted (item \#4 in our proposed roadmap). One challenge to consider may be the extent to which it is possible (or desirable) to develop general models that are fair across varied demographics versus creating personalized models for particular user groups. For example, success in developing accurate ASR for deaf speech has thus far focused on custom models for particular users \cite{GoogleSpeechImpairment, PowerPointCaption}. The need for personalization may be high given the ``long tail'' of disability, though the need to train personalized models may present additional barriers for PWD, as well as creating a two-tiered system of people for whom general AI models work by default and those for whom they do not. Involvement of PWD not only in evaluating AI systems, but also in defining meaningful usage scenarios, error metrics, and policies, is critical for the development of fair AI.

\section{Conclusion}
In this position paper, we have reflected on the ways in which current classes of AI systems, as well as several techniques that are the building blocks of AI, may limit the efficacy and fairness of these systems for people with disabilities. Ultimately, our goal is the creation of new design guidelines, datasets, algorithmic techniques, and error metrics that can help AI systems realize their enormous potential to benefit PWD, while avoiding the possible pitfalls we have outlined here. We hope this paper provides a research roadmap that can guide AI researchers and practitioners in creating systems that are fair to and effective for PWD.

\balance{}
\bibliographystyle{SIGCHI-Reference-Format}
\bibliography{main}


\begin{thebibliography}{00}


\ifx \showCODEN    \undefined \def \showCODEN     #1{\unskip}     \fi
\ifx \showDOI      \undefined \def \showDOI       #1{{\tt DOI:}\penalty0{#1}\ }
  \fi
\ifx \showISBNx    \undefined \def \showISBNx     #1{\unskip}     \fi
\ifx \showISBNxiii \undefined \def \showISBNxiii  #1{\unskip}     \fi
\ifx \showISSN     \undefined \def \showISSN      #1{\unskip}     \fi
\ifx \showLCCN     \undefined \def \showLCCN      #1{\unskip}     \fi
\ifx \shownote     \undefined \def \shownote      #1{#1}          \fi
\ifx \showarticletitle \undefined \def \showarticletitle #1{#1}   \fi
\ifx \showURL      \undefined \def \showURL       #1{#1}          \fi

\bibitem{agarwal2018reductions}
{Alekh Agarwal}, {Alina Beygelzimer}, {Miroslav Dud{\'\i}k}, {John Langford},
  {and} {Hanna Wallach}. 2018.
\newblock \showarticletitle{A reductions approach to fair classification}.
\newblock {\em arXiv preprint arXiv:1803.02453\/} (2018).
\newblock


\bibitem{MachineBias}
{Julia Angwin}, {Jeff Larson}, {Surya Mattu}, {Lauren Kirchner}, {and}
  {ProPublica}. 2016.
\newblock Machine Bias: There's software used across the country to predict
  future criminals. And it's biased against blacks.
\newblock   (2016).
\newblock
\showURL{%
Retrieved July 3, 2019 from
  \url{https://www.propublica.org/article/machine-bias-risk-assessments-in-criminal-sentencing}}


\bibitem{FaceID}
{{Apple Inc.}} 2019a.
\newblock About Face ID advanced technology.
\newblock   (2019).
\newblock
\showURL{%
Retrieved July 3, 2019 from \url{https://support.apple.com/en-us/HT208108}}


\bibitem{arkitbody}
{{Apple Inc.}} 2019b.
\newblock Augmented Reality -- ARKit 3 -- Apple Developer.
\newblock   (2019).
\newblock
\showURL{%
Retrieved July 3, 2019 from
  \url{https://developer.apple.com/augmented-reality/arkit/}}


\bibitem{AppleVoiceControl}
{{Apple Inc.}} 2019c.
\newblock macOS Catalina -- Introducing Voice Control. Your all-access to all
  devices.
\newblock   (2019).
\newblock
\showURL{%
Retrieved July 3, 2019 from
  \url{https://www.apple.com/macos/catalina-preview/\#accessibility}}


\bibitem{earlyadopters}
{Jeffrey~P. Bigham} {and} {Patrick Carrington}. 2018.
\newblock \showarticletitle{Learning from the Front: People with Disabilities
  as Early Adopters of AI}. In {\em HCIC 2018}.
\newblock


\bibitem{bolukbasi2016man}
{Tolga Bolukbasi}, {Kai-Wei Chang}, {James~Y Zou}, {Venkatesh Saligrama}, {and}
  {Adam~T Kalai}. 2016.
\newblock \showarticletitle{Man is to computer programmer as woman is to
  homemaker? debiasing word embeddings}. In {\em Advances in neural information
  processing systems}. 4349--4357.
\newblock


\bibitem{bragg2019sign}
{Danielle Bragg}, {Oscar Koller}, {Mary Bellard}, {Larwan Berke}, {Patrick
  Boudreault}, {Annelies Braffort}, {Naomi Caselli}, {Matt Huenerfauth},
  {Hernisa Kacorrim}, {Tessa Verhoef}, {Christian Vogler}, {and}
  {Meredith~Ringel Morris}. 2019.
\newblock \showarticletitle{Sign Language Recognition, Generation, and
  Translation: An Interdisciplinary Perspective}. In {\em Proceedings of the
  21st International ACM SIGACCESS Conference on Computers and Accessibility}.
  ACM.
\newblock


\bibitem{gendershades}
{Joy Buolamwini} {and} {Timnit Gebru}. 2018.
\newblock \showarticletitle{Gender shades: Intersectional accuracy disparities
  in commercial gender classification}. In {\em Conference on Fairness,
  Accountability and Transparency}. 77--91.
\newblock


\bibitem{chu2010providing}
{Min Chu}, {Yong Zhao}, {and} {Sheng Zhao}. 2010.
\newblock Providing personalized voice font for text-to-speech applications.
\newblock   (April~6 2010).
\newblock
\newblock
\shownote{US Patent 7,693,719.}


\bibitem{deepfake}
{{CNN Business}}. 2019.
\newblock Deepfake videos: Inside the Pentagon's race against disinformation.
\newblock   (2019).
\newblock
\showURL{%
Retrieved July 3, 2019 from
  \url{https://www.cnn.com/interactive/2019/01/business/pentagons-race-against-deepfakes/}}


\bibitem{deng2009imagenet}
{Jia Deng}, {Wei Dong}, {Richard Socher}, {Li-Jia Li}, {Kai Li}, {and} {Li
  Fei-Fei}. 2009.
\newblock \showarticletitle{Imagenet: A large-scale hierarchical image
  database}. In {\em 2009 IEEE conference on computer vision and pattern
  recognition}. Ieee, 248--255.
\newblock


\bibitem{duolingochatbot}
{Duolingo}. 2019.
\newblock Say hello to the Bots. The most advanced way to learn a language.
\newblock   (2019).
\newblock
\showURL{%
Retrieved July 3, 2019 from \url{http://bots.duolingo.com}}


\bibitem{uberaccident}
{The Economist}. 2018.
\newblock Why Uber's self-driving car killed a pedestrian.
\newblock   (2018).
\newblock
\showURL{%
Retrieved July 3, 2019 from
  \url{https://www.economist.com/the-economist-explains/2018/05/29/why-ubers-self-driving-car-killed-a-pedestrian}}


\bibitem{elias2010multi}
{John~Greer Elias}, {Wayne~Carl Westerman}, {and} {Myra~Mary Haggerty}. 2010.
\newblock Multi-touch gesture dictionary.
\newblock   (Nov.~23 2010).
\newblock
\newblock
\shownote{US Patent 7,840,912.}


\bibitem{GoogleSpeechImpairment}
{Engadget}. 2019.
\newblock Google trains its AI to accommodate speech impairments.
\newblock   (2019).
\newblock
\showURL{%
Retrieved July 3, 2019 from
  \url{https://www.engadget.com/2019/05/07/google-ai-impaired-speech-recognition/}}


\bibitem{objectdetectionbias}
{{Facebook Artificial Intelligence}}. 2019.
\newblock Does object recognition work for everyone? A new method to assess
  bias in CV systems.
\newblock   (2019).
\newblock
\showURL{%
Retrieved July 3, 2019 from
  \url{https://ai.facebook.com/blog/new-way-to-assess-ai-bias-in-object-recognition-systems/}}


\bibitem{feldman2015certifying}
{Michael Feldman}, {Sorelle~A Friedler}, {John Moeller}, {Carlos Scheidegger},
  {and} {Suresh Venkatasubramanian}. 2015.
\newblock \showarticletitle{Certifying and removing disparate impact}. In {\em
  Proceedings of the 21th ACM SIGKDD International Conference on Knowledge
  Discovery and Data Mining}. ACM, 259--268.
\newblock


\bibitem{fiannaca2018voicesetting}
{Alexander~J. Fiannaca}, {Ann Paradiso}, {Jon Campbell}, {and} {Meredith~Ringel
  Morris}. 2018.
\newblock \showarticletitle{Voicesetting: Voice Authoring UIs for Improved
  Expressivity in Augmentative Communication}. In {\em Proceedings of the 2018
  CHI Conference on Human Factors in Computing Systems} {\em (CHI '18)}. ACM,
  New York, NY, USA, Article 283, 12 pages.
\newblock
\showISBNx{978-1-4503-5620-6}
\showDOI{%
\url{http://dx.doi.org/10.1145/3173574.3173857}}


\bibitem{fok2018towards}
{Raymond Fok}, {Harmanpreet Kaur}, {Skanda Palani}, {Martez~E. Mott}, {and}
  {Walter~S. Lasecki}. 2018.
\newblock \showarticletitle{Towards More Robust Speech Interactions for Deaf
  and Hard of Hearing Users}. In {\em Proceedings of the 20th International ACM
  SIGACCESS Conference on Computers and Accessibility} {\em (ASSETS '18)}. ACM,
  New York, NY, USA, 57--67.
\newblock
\showISBNx{978-1-4503-5650-3}
\showDOI{%
\url{http://dx.doi.org/10.1145/3234695.3236343}}


\bibitem{SmartGate}
{Australian~Border Force}. 2019.
\newblock SmartGates.
\newblock   (2019).
\newblock
\showURL{%
Retrieved July 3, 2019 from
  \url{https://www.abf.gov.au/entering-and-leaving-australia/smartgates}}


\bibitem{fulmer2018using}
{Russell Fulmer}, {Angela Joerin}, {Breanna Gentile}, {Lysanne Lakerink}, {and}
  {Michiel Rauws}. 2018.
\newblock \showarticletitle{Using Psychological Artificial Intelligence (Tess)
  to Relieve Symptoms of Depression and Anxiety: Randomized Controlled Trial}.
\newblock {\em JMIR mental health\/} {5}, 4 (2018), e64.
\newblock


\bibitem{GooglePhotos}
{Google}. 2019a.
\newblock Google Photos Help -- Search by people, things, \& places in your
  photos.
\newblock   (2019).
\newblock
\showURL{%
Retrieved July 3, 2019 from
  \url{https://support.google.com/photos/answer/6128838?co=GENIE.Platform\%3DAndroid\&hl=en}}


\bibitem{GoogleAutomaticCaptioning}
{Google}. 2019b.
\newblock TouTube Help -- Use automatic captioning.
\newblock   (2019).
\newblock
\showURL{%
Retrieved July 3, 2019 from
  \url{https://support.google.com/youtube/answer/6373554?hl=en}}


\bibitem{lookout}
{Google}. 2019c.
\newblock With Lookout, discover your surroundings with the help of AI.
\newblock   (2019).
\newblock
\showURL{%
Retrieved July 3, 2019 from
  \url{https://www.blog.google/outreach-initiatives/accessibility/lookout-discover-your-surroundings-help-ai/}}


\bibitem{google-assistant-multiuser}
{{Google Assistant}}. 2017.
\newblock Tomato, tomahto. Google Home now supports multiple users.
\newblock   (2017).
\newblock
\showURL{%
Retrieved July 3, 2019 from
  \url{https://blog.google/products/assistant/tomato-tomahto-google-home-now-supports-multiple-users/}}


\bibitem{gottermeier2016user}
{Linda~G Gottermeier} {and} {S~Kushalnagar Raja}. 2016.
\newblock \showarticletitle{User Evaluation of Automatic Speech Recognition
  Systems for Deaf-Hearing Interactions at School and Work}.
\newblock {\em Audiology Today\/} {28}, 2 (2016), 20--34.
\newblock


\bibitem{grammarly}
{{Grammarly Inc.}} 2019.
\newblock Great Writing, Simplified.
\newblock   (2019).
\newblock
\showURL{%
Retrieved July 3, 2019 from \url{https://www.grammarly.com}}


\bibitem{guo2019statelens}
{Anhong Guo}, {Junhan Kong}, {Michael Rivera}, {Frank~F. Xu}, {and} {Jeffrey~P.
  Bigham}. 2019.
\newblock \showarticletitle{StateLens: A Reverse Engineering Solution for
  Making Existing Dynamic Touchscreens Accessible}. In {\em Proceedings of the
  32th Annual Symposium on User Interface Software and Technology} {\em (UIST
  '19)}. ACM, New York, NY, USA.
\newblock
\showDOI{%
\url{http://dx.doi.org/10.1145/3332165.3347873}}


\bibitem{guo2016tilt}
{Anhong Guo} {and} {Tim Paek}. 2016.
\newblock \showarticletitle{Exploring Tilt for No-touch, Wrist-only
  Interactions on Smartwatches}. In {\em Proceedings of the 18th International
  Conference on Human-Computer Interaction with Mobile Devices and Services}
  {\em (MobileHCI '16)}. ACM, New York, NY, USA, 17--28.
\newblock
\showISBNx{978-1-4503-4408-1}
\showDOI{%
\url{http://dx.doi.org/10.1145/2935334.2935345}}


\bibitem{vizwiz-dataset}
{Danna Gurari}, {Qing Li}, {Abigale~J. Stangl}, {Anhong Guo}, {Chi Lin},
  {Kristen Grauman}, {Jiebo Luo}, {and} {Jeffrey~P. Bigham}. 2018.
\newblock \showarticletitle{VizWiz Grand Challenge: Answering Visual Questions
  from Blind People}. In {\em Proceedings of the IEEE Conference on Computer
  Vision and Pattern Recognition}. 3608--3617.
\newblock


\bibitem{a11ywriting}
{Vicki~L. Hanson}, {Anna Cavender}, {and} {Shari Trewin}. 2015.
\newblock \showarticletitle{Writing About Accessibility}.
\newblock {\em Interactions\/} {22}, 6 (Oct. 2015), 62--65.
\newblock
\showISSN{1072-5520}
\showDOI{%
\url{http://dx.doi.org/10.1145/2828432}}


\bibitem{hardt2016equality}
{Moritz Hardt}, {Eric Price}, {Nati Srebro}, {and} {others}. 2016.
\newblock \showarticletitle{Equality of opportunity in supervised learning}. In
  {\em Advances in neural information processing systems}. 3315--3323.
\newblock


\bibitem{HireVue}
{HireVue}. 2019.
\newblock HireVue - Hiring Intelligence | Assessment \& Video Interview
  Software.
\newblock   (2019).
\newblock
\showURL{%
Retrieved July 3, 2019 from \url{https://www.hirevue.com}}


\bibitem{ARCaptioning}
{Dhruv Jain}, {Bonnie Chinh}, {Leah Findlater}, {Raja Kushalnagar}, {and} {Jon
  Froehlich}. 2018.
\newblock \showarticletitle{Exploring Augmented Reality Approaches to Real-Time
  Captioning: A Preliminary Autoethnographic Study}. In {\em Proceedings of the
  2018 ACM Conference Companion Publication on Designing Interactive Systems}
  {\em (DIS '18 Companion)}. ACM, New York, NY, USA, 7--11.
\newblock
\showISBNx{978-1-4503-5631-2}
\showDOI{%
\url{http://dx.doi.org/10.1145/3197391.3205404}}


\bibitem{jain2015dhh}
{Dhruv Jain}, {Leah Findlater}, {Jamie Gilkeson}, {Benjamin Holland}, {Ramani
  Duraiswami}, {Dmitry Zotkin}, {Christian Vogler}, {and} {Jon~E. Froehlich}.
  2015.
\newblock \showarticletitle{Head-Mounted Display Visualizations to Support
  Sound Awareness for the Deaf and Hard of Hearing}. In {\em Proceedings of the
  33rd Annual ACM Conference on Human Factors in Computing Systems} {\em (CHI
  '15)}. ACM, New York, NY, USA, 241--250.
\newblock
\showISBNx{978-1-4503-3145-6}
\showDOI{%
\url{http://dx.doi.org/10.1145/2702123.2702393}}


\bibitem{kamishima2012fairness}
{Toshihiro Kamishima}, {Shotaro Akaho}, {Hideki Asoh}, {and} {Jun Sakuma}.
  2012.
\newblock \showarticletitle{Fairness-aware classifier with prejudice remover
  regularizer}. In {\em Joint European Conference on Machine Learning and
  Knowledge Discovery in Databases}. Springer, 35--50.
\newblock


\bibitem{kane2017aac}
{Shaun~K. Kane}, {Meredith~Ringel Morris}, {Ann Paradiso}, {and} {Jon
  Campbell}. 2017.
\newblock \showarticletitle{``At Times Avuncular and Cantankerous, with the
  Reflexes of a Mongoose'': Understanding Self-Expression Through Augmentative
  and Alternative Communication Devices}. In {\em Proceedings of the 2017 ACM
  Conference on Computer Supported Cooperative Work and Social Computing} {\em
  (CSCW '17)}. ACM, New York, NY, USA, 1166--1179.
\newblock
\showISBNx{978-1-4503-4335-0}
\showDOI{%
\url{http://dx.doi.org/10.1145/2998181.2998284}}


\bibitem{kay2015unequal}
{Matthew Kay}, {Cynthia Matuszek}, {and} {Sean~A. Munson}. 2015.
\newblock \showarticletitle{Unequal Representation and Gender Stereotypes in
  Image Search Results for Occupations}. In {\em Proceedings of the 33rd Annual
  ACM Conference on Human Factors in Computing Systems} {\em (CHI '15)}. ACM,
  New York, NY, USA, 3819--3828.
\newblock
\showISBNx{978-1-4503-3145-6}
\showDOI{%
\url{http://dx.doi.org/10.1145/2702123.2702520}}


\bibitem{kim2012digits}
{David Kim}, {Otmar Hilliges}, {Shahram Izadi}, {Alex~D. Butler}, {Jiawen
  Chen}, {Iason Oikonomidis}, {and} {Patrick Olivier}. 2012.
\newblock \showarticletitle{Digits: Freehand 3D Interactions Anywhere Using a
  Wrist-worn Gloveless Sensor}. In {\em Proceedings of the 25th Annual ACM
  Symposium on User Interface Software and Technology} {\em (UIST '12)}. ACM,
  New York, NY, USA, 167--176.
\newblock
\showISBNx{978-1-4503-1580-7}
\showDOI{%
\url{http://dx.doi.org/10.1145/2380116.2380139}}


\bibitem{lakkaraju2017identifying}
{Himabindu Lakkaraju}, {Ece Kamar}, {Rich Caruana}, {and} {Eric Horvitz}. 2017.
\newblock \showarticletitle{Identifying unknown unknowns in the open world:
  Representations and policies for guided exploration}. In {\em Thirty-First
  AAAI Conference on Artificial Intelligence}.
\newblock


\bibitem{laput2016viband}
{Gierad Laput}, {Robert Xiao}, {and} {Chris Harrison}. 2016.
\newblock \showarticletitle{ViBand: High-Fidelity Bio-Acoustic Sensing Using
  Commodity Smartwatch Accelerometers}. In {\em Proceedings of the 29th Annual
  Symposium on User Interface Software and Technology} {\em (UIST '16)}. ACM,
  New York, NY, USA, 321--333.
\newblock
\showISBNx{978-1-4503-4189-9}
\showDOI{%
\url{http://dx.doi.org/10.1145/2984511.2984582}}


\bibitem{lewis2005hci}
{Clayton Lewis}. 2005.
\newblock \showarticletitle{HCI for people with cognitive disabilities}.
\newblock {\em ACM SIGACCESS Accessibility and Computing\/} 83 (2005), 12--17.
\newblock


\bibitem{moneyreader}
{LookTel}. 2019.
\newblock LookTel Money Reader.
\newblock   (2019).
\newblock
\showURL{%
Retrieved July 3, 2019 from \url{http://www.looktel.com/moneyreader}}


\bibitem{macleod2017caption}
{Haley MacLeod}, {Cynthia~L. Bennett}, {Meredith~Ringel Morris}, {and} {Edward
  Cutrell}. 2017.
\newblock \showarticletitle{Understanding Blind People's Experiences with
  Computer-Generated Captions of Social Media Images}. In {\em Proceedings of
  the 2017 CHI Conference on Human Factors in Computing Systems} {\em (CHI
  '17)}. ACM, New York, NY, USA, 5988--5999.
\newblock
\showISBNx{978-1-4503-4655-9}
\showDOI{%
\url{http://dx.doi.org/10.1145/3025453.3025814}}


\bibitem{taychatbot}
{Microsoft}. 2016.
\newblock Learning from Tay's introduction.
\newblock   (2016).
\newblock
\showURL{%
Retrieved July 3, 2019 from
  \url{https://blogs.microsoft.com/blog/2016/03/25/learning-tays-introduction/}}


\bibitem{kinectbody}
{Microsoft}. 2019a.
\newblock Azure Kinect DK -- BODY TRACKING SDK.
\newblock   (2019).
\newblock
\showURL{%
Retrieved July 3, 2019 from
  \url{https://azure.microsoft.com/en-us/services/kinect-dk/}}


\bibitem{voicefont}
{Microsoft}. 2019b.
\newblock Custom Voice.
\newblock   (2019).
\newblock
\showURL{%
Retrieved July 3, 2019 from \url{https://speech.microsoft.com/customvoice}}


\bibitem{MRGestures}
{Microsoft}. 2019c.
\newblock Gestures -- Mixed Reaity.
\newblock   (2019).
\newblock
\showURL{%
Retrieved July 3, 2019 from
  \url{https://docs.microsoft.com/en-us/windows/mixed-reality/gestures}}


\bibitem{immersivereader}
{Microsoft}. 2019d.
\newblock Immersive Reader -- An AI Service that helps users read and
  comprehend text.
\newblock   (2019).
\newblock
\showURL{%
Retrieved July 3, 2019 from
  \url{https://azure.microsoft.com/en-us/services/cognitive-services/immersive-reader/}}


\bibitem{seeingai}
{Microsoft}. 2019e.
\newblock Seeing AI.
\newblock   (2019).
\newblock
\showURL{%
Retrieved July 3, 2019 from \url{https://www.microsoft.com/en-us/ai/seeing-ai}}


\bibitem{WindowsHello}
{Microsoft}. 2019f.
\newblock Windows Hello: Discover facial recognition on Windows 10.
\newblock   (2019).
\newblock
\showURL{%
Retrieved July 3, 2019 from
  \url{https://www.microsoft.com/en-us/windows/windows-hello}}


\bibitem{morris2018dyslexia}
{Meredith~Ringel Morris}, {Adam Fourney}, {Abdullah Ali}, {and} {Laura
  Vonessen}. 2018.
\newblock \showarticletitle{Understanding the Needs of Searchers with
  Dyslexia}. In {\em Proceedings of the 2018 CHI Conference on Human Factors in
  Computing Systems} {\em (CHI '18)}. ACM, New York, NY, USA, Article 35, 12
  pages.
\newblock
\showISBNx{978-1-4503-5620-6}
\showDOI{%
\url{http://dx.doi.org/10.1145/3173574.3173609}}


\bibitem{motionanalysis}
{{Motion Analysis}}. 2019.
\newblock Sports Biomechanics.
\newblock   (2019).
\newblock
\showURL{%
Retrieved July 3, 2019 from
  \url{https://motionanalysis.com/industry/sports-biomechanics/}}


\bibitem{mott2018photography}
{Martez~E. Mott}, {Jane E.}, {Cynthia~L. Bennett}, {Edward Cutrell}, {and}
  {Meredith~Ringel Morris}. 2018.
\newblock \showarticletitle{Understanding the Accessibility of Smartphone
  Photography for People with Motor Impairments}. In {\em Proceedings of the
  2018 CHI Conference on Human Factors in Computing Systems} {\em (CHI '18)}.
  ACM, New York, NY, USA, Article 520, 12 pages.
\newblock
\showISBNx{978-1-4503-5620-6}
\showDOI{%
\url{http://dx.doi.org/10.1145/3173574.3174094}}


\bibitem{mott2016smarttouch}
{Martez~E. Mott}, {Radu-Daniel Vatavu}, {Shaun~K. Kane}, {and} {Jacob~O.
  Wobbrock}. 2016.
\newblock \showarticletitle{Smart Touch: Improving Touch Accuracy for People
  with Motor Impairments with Template Matching}. In {\em Proceedings of the
  2016 CHI Conference on Human Factors in Computing Systems} {\em (CHI '16)}.
  ACM, New York, NY, USA, 1934--1946.
\newblock
\showISBNx{978-1-4503-3362-7}
\showDOI{%
\url{http://dx.doi.org/10.1145/2858036.2858390}}


\bibitem{mott2019clustertouch}
{Martez~E. Mott} {and} {Jacob~O. Wobbrock}. 2019.
\newblock \showarticletitle{Cluster Touch: Improving Touch Accuracy on
  Smartphones for People with Motor and Situational Impairments}. In {\em
  Proceedings of the 2019 CHI Conference on Human Factors in Computing Systems}
  {\em (CHI '19)}. ACM, New York, NY, USA, Article 27, 14 pages.
\newblock
\showISBNx{978-1-4503-5970-2}
\showDOI{%
\url{http://dx.doi.org/10.1145/3290605.3300257}}


\bibitem{nicol2002children}
{Antony Nicol}, {Chris Casey}, {and} {Stuart MacFarlane}. 2002.
\newblock \showarticletitle{Children are ready for speech technology-but is the
  technology ready for them}.
\newblock {\em Interaction Design and Children, Eindhoven, The Netherlands\/}
  (2002).
\newblock


\bibitem{nuance-voice-biometric}
{{Nuance Communications, Inc.}} 2019.
\newblock Every voice matters: Our system knows who is talking and why.
\newblock   (2019).
\newblock
\showURL{%
Retrieved July 3, 2019 from
  \url{https://www.nuance.com/automotive/voice-biometrics.html}}


\bibitem{nushi2018towards}
{Besmira Nushi}, {Ece Kamar}, {and} {Eric Horvitz}. 2018.
\newblock \showarticletitle{Towards accountable AI: Hybrid human-machine
  analyses for characterizing system failure}. In {\em Sixth AAAI Conference on
  Human Computation and Crowdsourcing}.
\newblock


\bibitem{FBI-NGI}
{Federal~Bureau of Investigation}. 2019.
\newblock Next Generation Identification (NGI).
\newblock   (2019).
\newblock
\showURL{%
Retrieved July 3, 2019 from
  \url{https://www.fbi.gov/services/cjis/fingerprints-and-other-biometrics/ngi}}


\bibitem{oliver2013social}
{Mike Oliver}. 2013.
\newblock \showarticletitle{The social model of disability: Thirty years on}.
\newblock {\em Disability \& society\/} {28}, 7 (2013), 1024--1026.
\newblock


\bibitem{orcam}
{OrCam}. 2019.
\newblock OrCam MyEye 2 -- For the Blind and Visually Impaired.
\newblock   (2019).
\newblock
\showURL{%
Retrieved July 3, 2019 from \url{https://www.orcam.com/en/myeye2/}}


\bibitem{knfbreader}
{KNFB Reader}. 2018.
\newblock KNFB Reader gives you easy access to print and files, anytime,
  anywhere.
\newblock   (2018).
\newblock
\showURL{%
Retrieved July 3, 2019 from \url{https://knfbreader.com}}


\bibitem{rello2015spellchecker}
{Luz Rello}, {Miguel Ballesteros}, {and} {Jeffrey~P. Bigham}. 2015.
\newblock \showarticletitle{A Spellchecker for Dyslexia}. In {\em Proceedings
  of the 17th International ACM SIGACCESS Conference on Computers \&\#38;
  Accessibility} {\em (ASSETS '15)}. ACM, New York, NY, USA, 39--47.
\newblock
\showISBNx{978-1-4503-3400-6}
\showDOI{%
\url{http://dx.doi.org/10.1145/2700648.2809850}}


\bibitem{rodger2004field}
{James~A Rodger} {and} {Parag~C Pendharkar}. 2004.
\newblock \showarticletitle{A field study of the impact of gender and user's
  technical experience on the performance of voice-activated medical tracking
  application}.
\newblock {\em International Journal of Human-Computer Studies\/} {60}, 5-6
  (2004), 529--544.
\newblock


\bibitem{schlogl2013seniors}
{S. Schl\"{o}gl}, {G. Chollet}, {M. Garschall}, {M. Tscheligi}, {and} {G.
  Legouverneur}. 2013.
\newblock \showarticletitle{Exploring Voice User Interfaces for Seniors}. In
  {\em Proceedings of the 6th International Conference on PErvasive
  Technologies Related to Assistive Environments} {\em (PETRA '13)}. ACM, New
  York, NY, USA, Article 52, 2 pages.
\newblock
\showISBNx{978-1-4503-1973-7}
\showDOI{%
\url{http://dx.doi.org/10.1145/2504335.2504391}}


\bibitem{tatman2017gender}
{Rachael Tatman}. 2017.
\newblock \showarticletitle{Gender and dialect bias in YouTube's automatic
  captions}. In {\em Proceedings of the First ACL Workshop on Ethics in Natural
  Language Processing}. 53--59.
\newblock


\bibitem{facebookchatbots}
{TechCrunch}. 2016.
\newblock Facebook launches Messenger platform with chatbots.
\newblock   (2016).
\newblock
\showURL{%
Retrieved July 3, 2019 from
  \url{https://techcrunch.com/2016/04/12/agents-on-messenger/}}


\bibitem{FacebookHousingDiscrimination}
{{The New York Times}}. 2019.
\newblock Facebook Engages in Housing Discrimination With Its Ad Practices,
  U.S. Says.
\newblock   (2019).
\newblock
\showURL{%
Retrieved July 3, 2019 from
  \url{https://www.nytimes.com/2019/03/28/us/politics/facebook-housing-discrimination.html}}


\bibitem{personalitytests}
{TopResume}. 2019.
\newblock Get to Know the 5 Most Popular Pre-Employment Personality Tests.
\newblock   (2019).
\newblock
\showURL{%
Retrieved July 3, 2019 from
  \url{https://www.topresume.com/career-advice/how-to-pass-the-pre-employment-personality-test}}


\bibitem{tranter2006overview}
{Sue~E Tranter} {and} {Douglas~A Reynolds}. 2006.
\newblock \showarticletitle{An overview of automatic speaker diarization
  systems}.
\newblock {\em IEEE Transactions on audio, speech, and language processing\/}
  {14}, 5 (2006), 1557--1565.
\newblock


\bibitem{trewin2018ai}
{Shari Trewin}. 2018.
\newblock \showarticletitle{AI Fairness for People with Disabilities: Point of
  View}.
\newblock {\em arXiv preprint arXiv:1811.10670\/} (2018).
\newblock


\bibitem{umichpedestrian}
{{University of Michigan}}. 2019.
\newblock Teaching self-driving cars to predict pedestrian movement.
\newblock   (2019).
\newblock
\showURL{%
Retrieved July 3, 2019 from
  \url{https://news.umich.edu/teaching-self-driving-cars-to-predict-pedestrian-movement/}}


\bibitem{PowerPointCaption}
{VentureBeat}. 2019.
\newblock How Microsoft is using AI to improve accessibility.
\newblock   (2019).
\newblock
\showURL{%
Retrieved July 3, 2019 from
  \url{https://venturebeat.com/2019/05/06/how-microsoft-is-using-ai-to-improve-accessibility/}}


\bibitem{LiveCaption}
{The Verge}. 2019.
\newblock Android Q's Live Caption feature adds real-time subtitles to any
  audio or video playing on your phone.
\newblock   (2019).
\newblock
\showURL{%
Retrieved July 3, 2019 from
  \url{https://www.theverge.com/2019/5/7/18528447/google-android-q-live-caption-video-transcription-io-2019}}


\bibitem{vtyurina2019verse}
{Alexandra Vtyurina}, {Adam Fourney}, {Meredith~Ringel Morris}, {Leah
  Findlater}, {and} {Ryen White}. 2019.
\newblock \showarticletitle{VERSE: Bridging Screen Readers and Voice Assistants
  for Enhanced Eyes-Free Web Search}. In {\em Proceedings of the 21st
  International ACM SIGACCESS Conference on Computers and Accessibility}. ACM.
\newblock


\bibitem{wang2003silhouette}
{Liang Wang}, {Tieniu Tan}, {Huazhong Ning}, {and} {Weiming Hu}. 2003.
\newblock \showarticletitle{Silhouette analysis-based gait recognition for
  human identification}.
\newblock {\em IEEE transactions on pattern analysis and machine
  intelligence\/} {25}, 12 (2003), 1505--1518.
\newblock


\bibitem{zyskowski2015accessiblecrowdwork}
{Kathryn Zyskowski}, {Meredith~Ringel Morris}, {Jeffrey~P. Bigham}, {Mary~L.
  Gray}, {and} {Shaun~K. Kane}. 2015.
\newblock \showarticletitle{Accessible Crowdwork?: Understanding the Value in
  and Challenge of Microtask Employment for People with Disabilities}. In {\em
  Proceedings of the 18th ACM Conference on Computer Supported Cooperative Work
  \&\#38; Social Computing} {\em (CSCW '15)}. ACM, New York, NY, USA,
  1682--1693.
\newblock
\showISBNx{978-1-4503-2922-4}
\showDOI{%
\url{http://dx.doi.org/10.1145/2675133.2675158}}


\end{thebibliography}

\end{document}